\documentclass[a4paper,useAMS,usenatbib]{mnras}

\usepackage[T1]{fontenc}
\usepackage{ae,aecompl}

\usepackage{graphicx}	
\usepackage{amsmath}	
\usepackage{amssymb}	
\usepackage{hyperref}  
\usepackage{soul,array}  

\defcitealias{Garrison-Kimmel2014b}{GK14}

\newcommand{\bc}{\begin{center}}
\newcommand{\ec}{\end{center}}

\newcommand{\kpc}            {\,{\rm kpc}}
\newcommand{\Mpc}            {\,{\rm Mpc}}
\newcommand{\Msun}           {\,{\rm M}_\odot}

\newcommand{\kms}            {\,{\rm km}\,\,{\rm s}^{-1}}

\newcommand{\Mstr}          {M_{\rm str}}
\newcommand{\vmax}          {V_{\rm max}}

\newcommand{\Mout}          { {\rm M_{\rm 3Mpc}}}
\newcommand{\vr}            {V_{\rm r}}

\newcommand{\sigmalos}      {\sigma_{\rm los}}

\newcommand{\apostle}       {{\small APOSTLE }}
\newcommand{\aphydro}       {{\small AP-HYDRO }}
\newcommand{\apdmo}         {{\small AP-DMO }}
\newcommand{\dove}          {{\small DOVE }}
\newcommand{\subfind}       {{\small SUBFIND }}
\newcommand{\fig}           {Fig.~\ref}

\newcommand{\red}[1]{\textcolor{black}{#1}}

\title[LG missing dwarfs] {The missing dwarf galaxies of the Local Group}
\author[A. Fattahi et al. ]{Azadeh Fattahi$^1$\thanks{Email: azadeh.fattahi-savadjani@durham.ac.uk},
  Julio F. Navarro$^2$, and 
  Carlos S. Frenk$^1$
  \\ \\
  $^1$Institute for Computational Cosmology, Department of Physics,
  University of Durham, South Road, Durham DH1 3LE, UK\\
  $^2$CIfAR Fellow. Department of Physics and Astronomy,University of Victoria, PO Box
  3055 STN CSC, Victoria, BC, V8W 3P6, Canada\\
}

\date{Accepted XXX. Received YYY; in original form ZZZ}

\pubyear{2019}

\begin{document}
\label{firstpage}
\pagerange{\pageref{firstpage}--\pageref{lastpage}}
\maketitle

\begin{abstract}
We study the Local Group (LG) dwarf galaxy population
predicted by the \apostle $\Lambda$CDM cosmological hydrodynamics
simulations. These indicate that: (i)~the total mass within $3$ Mpc of
the Milky Way-Andromeda midpoint ($M_{\rm 3Mpc}$) typically exceeds
$\sim 3$ times the sum of the virial masses ($M_{\rm 200crit}$) of the
two primaries and (ii)~the dwarf galaxy formation efficiency per unit
mass is uniform throughout the volume. This suggests that the
satellite population within the virial radii of the Milky Way and
Andromeda should make up fewer than one third of all LG dwarfs
within $3$ Mpc. This is consistent with the fraction of observed LG
galaxies with stellar mass $M_*>10^7\Msun$ that are satellites
($12$ out of $42$; i.e., $28$ per cent). For the \apostle galaxy
mass-halo mass relation, the total number of such galaxies 
further suggests a LG mass of $M_{\rm 3 Mpc}\sim 10^{13}\Msun$. At
lower galaxy masses, however, the observed satellite fraction is
substantially higher ($42$ per cent for $M_*>10^5\Msun$). If this is due to incompleteness in the field sample,
then $\sim 50$ dwarf galaxies at least as massive as the Draco dwarf
spheroidal must be missing from the current LG {\it field} dwarf
inventory. The incompleteness interpretation is supported by the
pronounced flattening of the LG luminosity function below
$M_*\sim 10^7\Msun$, and by the scarcity of low-surface brightness LG
field galaxies compared to satellites. The simulations indicate that
most missing dwarfs should lie near the virial boundaries of the two
LG primaries, and predict a trove of nearby dwarfs that await
discovery by upcoming wide-field imaging surveys.

\end{abstract}

\begin{keywords}
Local Group -- galaxies: dwarf -- dark matter 
\end{keywords}

\section{Introduction}
\label{SecIntro}

The inventory of galaxies in the surroundings of the Milky~Way (MW)
and Andromeda (M31) galaxies is almost certainly incomplete. Every new
wide-field imaging survey of the night sky, when combined with refined
galaxy finding techniques, yields almost inevitably a plentiful catch
of new discoveries, many of which have been MW and M31
satellites \citep[see, e.g.,][]{Belokurov2007,Belokurov2010,
  McConnachie2008, Koposov2015,Drlica-Wagner2015,Bechtol2015}. In the
case of the MW, for example, satellites as faint as
$M_V\sim -2$ have now been reported, extending by several orders of
magnitude the faint-end limit of the galaxy luminosity function to a
``confusion-limited'' regime where faint galaxies and star clusters
become indistinguishable from each other without accurate kinematic
data
\citep{Simon2007,Tolstoy2009,McConnachie2012,Koposov2011,Martin2016b,Walker2016,Simon2019}.

These newly-discovered satellites are not just faint but have also
extremely low surface brightness (LSB), reaching values $\sim 100$
times fainter than the ``ultra diffuse'' galaxy population (UDGs)
recently reported in galaxy clusters and in the vicinity of bright
galaxies \citep{vanDokkum2015,Yagi2016}.  These extreme LSB dwarfs are
rather challenging to detect and, consequently, the census of faint
satellites around the MW and M31 is widely agreed to be far from
complete \citep{Newton2018,Nadler2019}. Correcting for this
incompleteness is non trivial, for it requires making assumptions
about how satellites populate the size-luminosity plane, as well as
how they are distributed radially, neither of which is known
accurately enough \citep{Koposov2008}.

The incompleteness is often thought to affect solely the
``ultra-faint'' regime, defined here as galaxies below $M_V\sim -8$
(or a stellar mass of $M_*\sim 10^5\Msun$; i.e., that of the Draco, or
Ursa Minor, dwarf spheroidals; hereafter dSphs). However, the recent
discovery of ``feeble giant'' galaxies, such as the Crater II and
Antlia II dSphs \citep{Torrealba2016,Torrealba2018}, suggests that our
current inventory of MW satellites may be missing systems even
in the $M_V<-8$ luminosity range typically referred to as ``classical
dSphs''. Galaxies like Crater II have unexpectedly large radii, making
them practically invisible through the foreground of Galactic stars
and the background of distant galaxies.

Although difficult to detect, these ultra-diffuse systems have the
potential of yielding important clues to our understanding of dwarf
galaxy formation. Indeed, the kinematics of their stars have proved
challenging to interpret: some have extremely low velocity dispersions,
hinting at very low dark matter densities \citep[e.g., Crater II, And
  XIX,][]{Collins2013,Caldwell2017}, while others inhabit surprisingly
dense haloes, according to the kinematical evidence \citep[e.g., Draco,
  Tucana,][]{Walker2007,Gregory2019}. These diverse properties are
mirrored by other UDGs outside the Local Group (LG), where some have been
associated with massive dark haloes \citep{vanDokkum2016,Beasley2016},
whereas others have been found to contain little obvious evidence for
dark matter \citep{vanDokkum2018}.

\begin{figure}
  \hspace{-0.3cm} \resizebox{8.5cm}{!}{\includegraphics{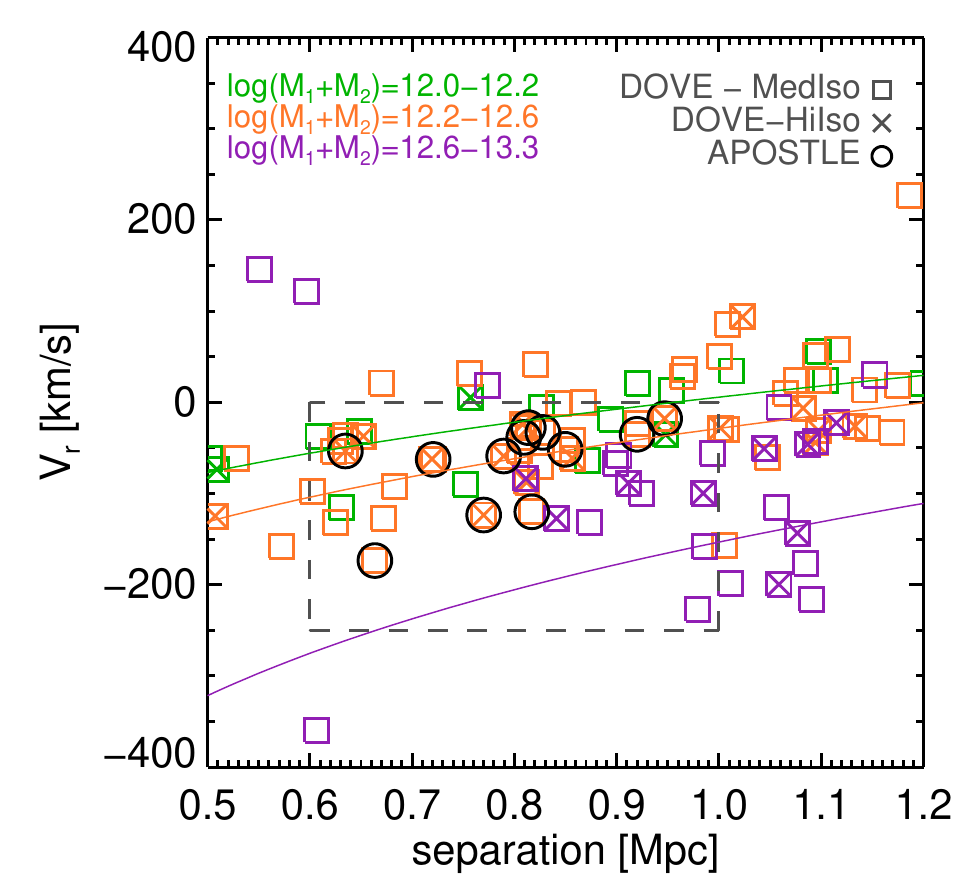}}\\%
  \caption{Relative radial velocity vs. separation of halo pairs in
    the DOVE cosmological simulation. Squares and crosses correspond
    to the medium-isolation (MedIso) and high-isolation (HiIso)
    criteria. \apostle candidates are highlighted with circles. Solid
    curves indicate the timing argument solutions of $\log
    (M_1+M_2)/\Msun=12.1$, $12.4$, and $13$ coloured in green, orange
    and purple, respectively. Similarly, DOVE pairs are colour coded
    in the same way, according to the combined virial mass of the
    pair, $M_1+M_2$. The box shows the radial velocity and separation
    criteria chosen to select LG-like pairs.}
\label{FigVrD}
\end{figure}

It would be extremely valuable to find other examples of such systems
in the Local Group, as their proximity facilitates their study. Particularly
interesting are luminous dwarfs such as the ``feeble
giants'' mentioned above, where the large number of giant
stars amenable to spectroscopic observation would enable detailed modelling
that may shed light on the origin of their puzzling kinematics.

We address these issues here using cosmological hydro dynamical
simulations of Local Group-like volumes from the APOSTLE\footnote{A
  Project Of Simulating The Local Environment}
project \citep{Fattahi2016,Sawala2016b}. In particular, we aim to
estimate the number of ``classical dSphs'' (i.e., those with $M_V<-8$,
or, equivalently, $M_*>10^5\Msun$) in the Local Group volume, defined
here as the $3$ Mpc-radius sphere around the midpoint between the MW and
M31. As we discuss below, this depends primarily on the total mass
within that volume, and on whether the dwarf galaxy formation
``efficiency'' (i.e., the number of galaxies per unit mass) in that
volume is substantially different from that of the MW halo.

The plan for this paper is as follows. After describing the
simulations (Sec.~\ref{sec:sim}) and galaxy samples (Sec.~\ref{SecLG})
we consider in our study, we analyse the galaxy formation efficiency
in the LG field and around the two primaries
(Sec~\ref{sec:demographics}).  The analysis contrasts the expected
faint-end of the LG luminosity/stellar mass function with current
observational constraints, and yields an estimate of the total mass
within $3$~Mpc (Sec.~\ref{sec:LGmass}); a prediction for the number of
luminous dwarfs missing from that volume
(Sec.~\ref{SecIncompleteness}); and clear indications as to where they
might be located (Sec.~\ref{sec:missingDwarfs}). We end with a brief
discussion of the surface brightness limits of current samples of
satellites and field LG galaxies, which clearly demonstrates the lack
of known low surface brightness galaxies in the LG field.

\section{Numerical Simulations}
\label{sec:sim}

\subsection{The DOVE simulation}
\label{sec:dove}

We use the \dove N-body cosmological simulations \citep{Jenkins2013},
to select LG-like environments. \dove followed evolution of a $100^3
\Mpc^3$ cosmological cube with $1620^3$ collisionless DM particles of
mass $m_p=8.8\times10^6 \Msun $. The simulation started at redshift
$z=127$ and was run to $z=0$ with the Tree-PM code {\small P-Gadget3},
a variant of the publicly available code, {\small Gadget-2}
\citep{Springel2005a}. \dove adopts flat $\Lambda$CDM cosmological
parameters consistent with WMAP-7 data \citep{Komatsu2011}: $h=0.704$,
$\Omega_{\rm m}=0.272$, $\Omega_{\rm bar}=0.045$.

DM haloes are identified in \dove at $z=0$ using the
friends-of-friends algorithm \citep[FoF;][]{Davis1985} with linking
length $0.2$ times the mean interparticle separation. Bound
structures and substructures in FoF haloes are found recursively
using \subfind \citep{Springel2001a}.

We search for LG-like environments in \dove by considering all haloes
with virial\footnote{We define the virial boundary of a halo as that
  of a sphere with mean interior density equal to $200$ times the
  critical density of the Universe.} mass above $5\times 10^{11}
\Msun$ and identifying pairs with separations $\sim 800 \kpc$. The
pair members are typically in separate FoF groups but in some cases
they are linked into a single FoF halo.

We select pairs that meet a relatively strict isolation criterion
(``MedIso'') so that there are no haloes more massive than the
lower-mass pair member within $d_{\rm iso}=2.5 \Mpc$ from the midpoint
between the primaries\footnote{We shall refer to the midpoint between
  the pair members as the ``barycentre'' for short.}. A more
restrictive isolation criterion (``HiIso'') was also considered, with
$d_{\rm iso}=5 \Mpc$.

\fig{FigVrD} shows the separation and relative radial velocity ($\vr$)
of \dove pairs. Systems identified with the ``MedIso'' and ``HiIso''
isolation criteria are represented with open squares and crosses,
respectively. (HiIso pairs are a subsample of the MedIso sample, by
definition). The pairs are colour-coded according to the sum of the
virial mass of the paired haloes, $M_1+M_2$. The
solid curves indicate the expected loci of pairs of fixed total mass
on radial orbits, according to the timing argument
\citep{Kahn1959,Li2008,Fattahi2016}.

We narrow down the pair selection by applying constraints on
separation, $600<r/\kpc<1000$, and radial velocity, $-250<\vr/\kms<0$,
respectively, in order to approximate the present-day kinematics of
the MW-M31 pair. Additionally, we impose a minimum mass ratio cut of
$M_2/M_1>0.3$ ($M_1>M_2$) to discard pairs with a large mass
difference. Hereafter, we shall use ``\dove pairs'' to refer to all
MedIso and HighIso pairs satisfying these conditions.

\subsection{\apostle simulations}
\label{sec:apostle}

The \apostle project is a suite of cosmological hydrodynamical
re-simulations of 12 volumes selected from the \dove sample of LG
candidate volumes discussed above \citep{Sawala2016b,Fattahi2016}.  In
addition to the kinematic constraints described in the \dove
selection, \apostle pair members satisfy a relative tangential
velocity criterion, $V_{\rm t}<100 \kms$, and the surrounding haloes
follow the Hubble flow out to $\sim 4 \Mpc$,
which is observed to be only weakly decelerated beyond $\sim 1$ Mpc
\citep[see][for details]{Fattahi2016}. \apostle volumes are a
subsample of the MedIso pairs described in the previous section, and
are highlighted with circles in \fig{FigVrD}.

The \apostle simulations were run at three different levels of
resolutions, labelled AP-L1, AP-L2, and AP-L3 with initial mass per gas
particle of $\sim10^4$, $10^5$, and $10^6 \Msun$, respectively, using
the code developed for the EAGLE project \citep{Schaye2015,Crain2015}.
All 12 \apostle volumes were simulated at resolution levels L2 and
L3, but only 5 volumes have been simulated at resolution L1, due to the
computational expense.

The EAGLE galaxy formation model was calibrated to reproduce the observed
$z=0.1$ stellar mass function of galaxies in the range $M_*\sim
10^{8}$-$10^{11} \Msun$, in a cosmological volume ($100^3$ Mpc$^3$), as
well as the stellar mass-size relation of galaxies. The
subgrid physics model includes metal cooling, star formation, stellar
evolution and supernovae feedback, a homogeneous background UV/X-Ray
photoionisation radiation, supermassive black hole formation and
evolution, and AGN feedback. The model reproduces the rotation curve
of galaxies quite well \citep{Schaller2015a}, as well as the
Tully-Fisher relation over a wide range of masses \citep{Ferrero2017,Sales2017}.

Haloes, subhaloes, and galaxies in \apostle are also identified using
the FoF and \subfind algorithms. Each \apostle volume has a
dark-matter-only (DMO) counterpart. We refer to the hydrodynamical and
dark-matter-only runs as \aphydro and \apdmo, respectively.

\section{Galaxy samples}
\label{SecLG}

\subsection{Simulated galaxy sample}

Galaxies and haloes in the simulated LG-like volumes are identified as
bound structures, found by \subfind within 3~Mpc from the pair
barycentre. We hereafter refer to the MW and M31 galaxy analogues as
``primaries''. Satellites are identified as galaxies (or subhaloes in
the case of DMO runs) located within the virial radius of each of the
primaries.

For the central system of each FoF halo we define a galaxy stellar
mass as that enclosed within a radius, $r_{\rm gal}=0.15\times$ the
virial radius of its halo. For subhaloes, where virial radii are
ill-defined, we use an average relation derived from the \apostle
centrals: $r_{\rm gal}=0.169 (V_{\rm max}/$km s$^{-1})^{1.01}$ kpc, where
$V_{\rm max}$ is the maximum circular velocity of the
system\footnote{The maximum circular velocity of a halo is a useful
  proxy of its virial mass. For isolated haloes, and at $z=0$, the
  tight relation between the two may be approximated as
  $M_{200}/(10^{10}\Msun) = 19 \, (V_{\rm max}/100 \kms$ $)^{3.04}$,
  based on \apostle centrals}.

\subsection{The Local Group galaxy sample}

Our main source of LG galaxies is the latest version of the LG
catalogue of \cite{McConnachie2012}\footnote{available from
  \url{www.astro.uvic.ca/~alan/Nearby_Dwarf_Database.html}}, with a
few updates with more recent discoveries (i.e. Crater~II and
Antlia~II). We extend the catalogue with galaxies from the online
Extragalactic Distance Database\footnote{\url{edd.ifa.hawaii.edu}}
\citep{Tully2009} with reliable distance measurements, i.e. from the
tip of the red giant branch or Cepheid variables methods\footnote{The
  majority of the distance measurements are based on
  \citet{Dalcanton2009}, with the rest from
  \citet{Saha2002,Karachentsev2006,Karachentsev2007,Makarova2005}.}. This
results in the addition of $\sim 10$ dwarf galaxies in the outskirts
of the Local Group.

Our analysis uses the position, distance, V-band apparent magnitude,
and half-light radius of a galaxy. We estimate stellar masses using
the V-band stellar mass-to-light ratios given in table 1 of
\citet{Woo2008} for individual galaxies, or use their table 2
otherwise. Some field galaxies only have B-band magnitudes, in which
case we use the B-band stellar mass-to-light ratios from
\citet{Woo2008}.

We define the Local Group volume as that of a $3$ Mpc sphere centred
at the midpoint between MW and M31\footnote{As for the simulated
  sample, we assume, for simplicity, that the ``LG barycentre''
  coincides with the midpoint between MW and
  M31.}. \red{Since $r_{200}$ is not well known for either the  MW
  and M31,} we consider as ``satellites'' any dwarf within $300$ kpc
from the centre of either MW or M31. We use the term ``Local Group
field'' galaxy to denote isolated galaxies (i.e, not MW or M31
satellites) within $3$ Mpc from the LG barycentre.

\begin{figure}
  \hspace{-0.2cm}
  \resizebox{8.6cm}{!}{\includegraphics{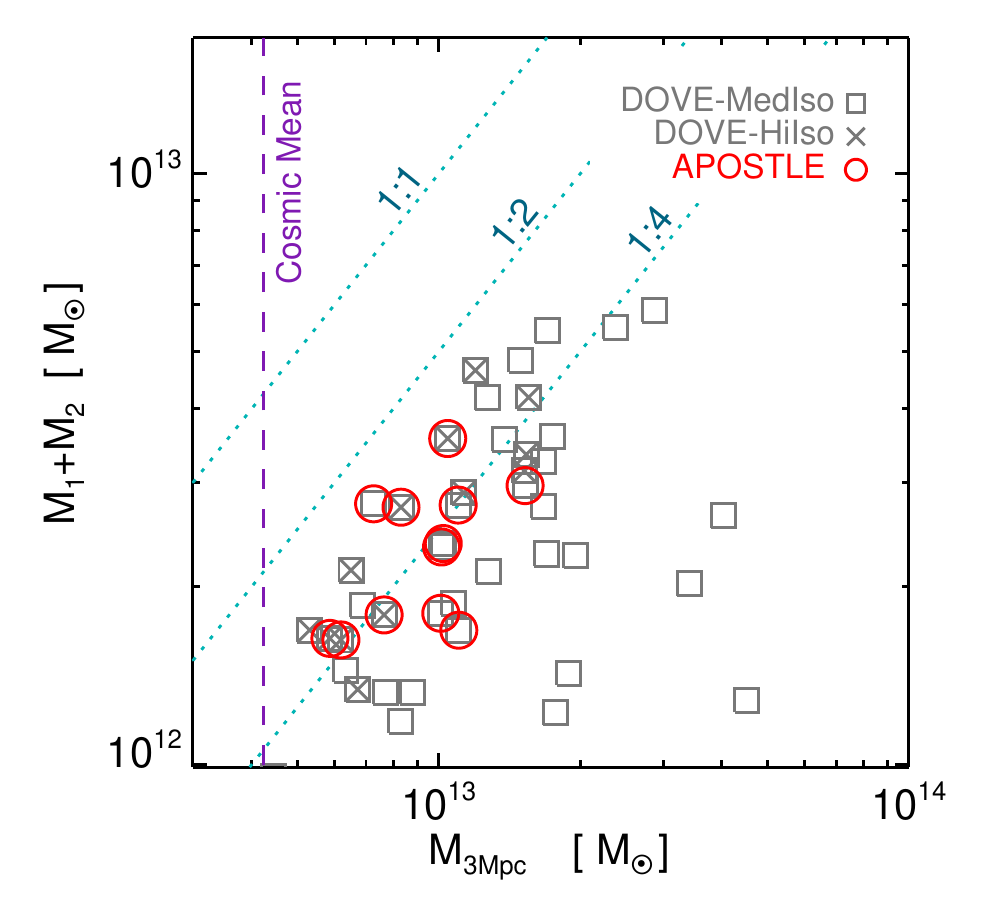}}\\%
  \caption{Combined virial mass of the pair, $M_1+M_2$, versus the
    total mass, $M_{\rm 3Mpc}$, within a sphere of $3$ Mpc radius
    centred on the barycentre, for MedIso and HiIso \dove pairs. The
    vertical dashed line indicates $M_{\rm 3Mpc}$ for a sphere with
    mean density equal to the mean matter density of the
    Universe. \red{The diagonal dotted lines show constant
      ratios between $M_1+M_2$ and $M_{\rm 3Mpc}$, as indicate by
      their labels.}. \apostle candidates are marked with red circles. }
\label{FigDove1}
\end{figure}

\begin{figure}
  \hspace{-0.2cm}
  \resizebox{8.4cm}{!}{\includegraphics{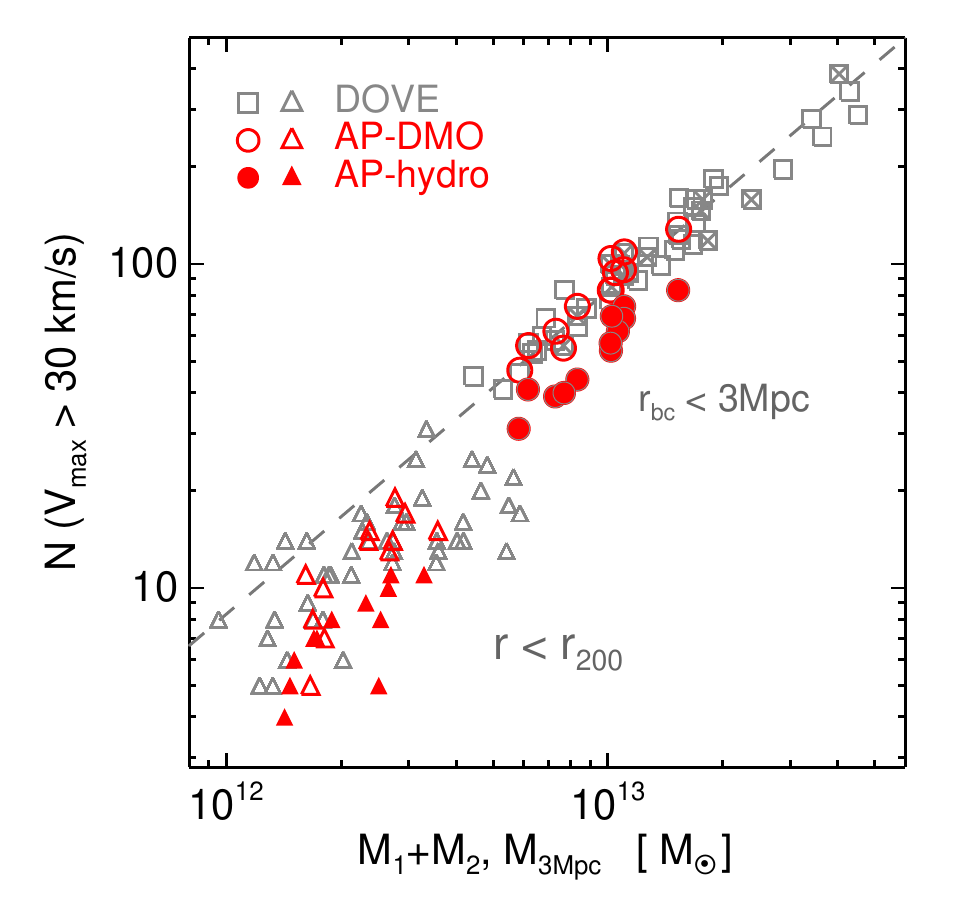}}\\%
  \caption{Total number of haloes/subhaloes with $\vmax>30 \kms$
    within a 3Mpc-radius sphere centred on the pair barycentre, versus
    the total mass within the same volume, $M_{\rm 3Mpc}$, for MedIso
    (grey squares) and HiIso (crosses) \dove pairs. Results for 12
    \apostle dark matter only (DMO) simulations and their
    hydrodynamical counterparts are shown as open and solid red
    circles, respectively. The triangles indicate subhaloes of the two
    primaries (satellites) vs the combined virial mass of the
    primaries.}
\label{FigDove2}
\end{figure}

\section{The demographics of LG-like environments}
\label{sec:demographics}

\subsection{Total mass within 3 Mpc}
\label{sec:mass}

We begin our analysis by considering the total mass of the LG-like
volumes selected from \dove. This is illustrated in \fig{FigDove1},
where we show, in the left panel, $M_{\rm 3Mpc}$, the total mass
within $3$ Mpc of each pair midpoint, versus the combined virial mass
of the pair, $M_1+M_2$. The vertical dashed line in \fig{FigDove1}
indicates the mass expected if the LG volumes had the same density as
the average matter density of the Universe. LG environments are
clearly overdense, and the overdensity increases systematically with
the combined mass of the pair.

All LG-like volumes lie below the 1:2 line in this plot, indicating
that there is at least as much mass around the primaries as in the
primaries themselves, and often much more. This is true even for the
highly isolated pairs (HighIso, identified with crosses in
\fig{FigDove1}), for which, on average,
$M_{\rm 3Mpc}\sim 3.7(M_1+M_2)$. For \apostle volumes, which contain
some MedIso and some HighIso volumes and are identified with red
circles in \fig{FigDove1}, the average $M_{\rm 3Mpc}/(M_1+M_2)$ is
$4.2$.

We note that the extra mass outside the primary haloes is not expected
to be distributed isotropically in the considered volume, but rather
in filaments and sheet-like structures \citep[see figure 2 of][for a
visual impression of mass distribution around \apostle
volumes]{Penarrubia2017}.

\subsection{Haloes and subhaloes}
\label{sec:subhaloes}

The higher the total LG mass, the larger the number of haloes (and,
hence, galaxies) that it is expected to contain. We show this
explicitly in \fig{FigDove2}, where the grey squares and circles
indicate the {\it total} number of haloes and subhaloes in DMO runs
with maximum circular velocity, $V_{\rm max}$, exceeding
$30\kms$. This velocity roughly corresponds to the minimum mass
expected of haloes that host dwarfs as luminous as the ``classical
dSphs'', i.e. those with $M_*>10^5\Msun$ \citep[e.g.][]{Guo2010,Moster2013,Behroozi2013c,Sawala2015,GarrisonKimmel2019}.

The grey squares in \fig{FigDove2} show
the tight relation between the total number of such haloes and the
total mass within 3 Mpc; indeed, the rms scatter around a 1:1 linear
fit is only $0.1$ dex. Filled red circles show the results for the
\aphydro runs, which also follow the 1:1 trend, but lie a factor of
$\sim 1.5$ below the DMO results. This is because low-mass haloes (which
dominate the total count) are affected by the loss of baryons driven
by photoionisation and supernova feedback. This loss stunts the mass
growth of the haloes, leading to lower values of $V_{\rm max}$ than
their DMO counterparts \citep{Sawala2012,Sawala2016a}.

Finally, the triangles in \fig{FigDove2} indicate the number of
subhaloes within the virial radii of the two primaries, as a function
of their host virial mass ($M_1+M_2$). Here, the effects of tidal
stripping tend to depress systematically the numbers below
extrapolations of the $1$:$1$ trend seen at larger masses, as well as
to increase the scatter. \red{We have explictly checked
  that the number of subhaloes
  does follow the same $1$:$1$ trend as the primaries when using their
  peak maximum circular velocity, which is typically reached just
  before infall into the primary and is thus unaffected by tidal
  stripping.}

\begin{figure}
  \hspace{-0.3cm}
  \resizebox{8.6cm}{!}{\includegraphics{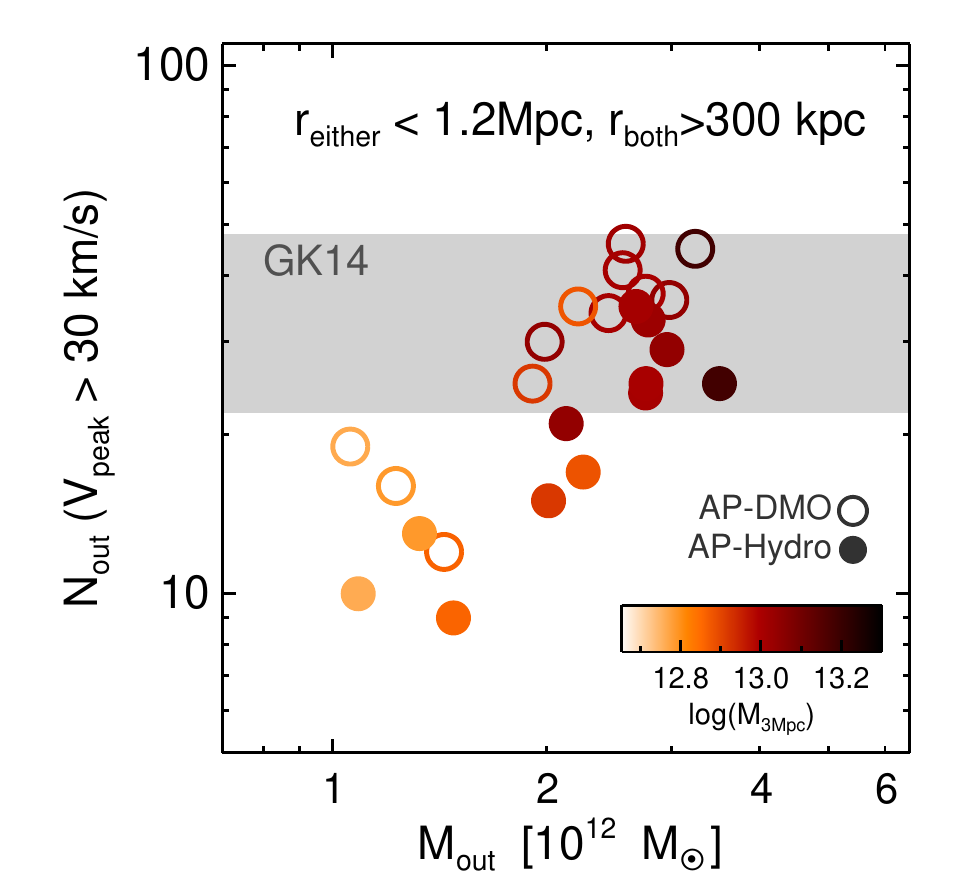}}\\%
  \caption{Number of haloes/subhaloes in AP-L2 simulations with
    $V_{\rm peak}$ exceeding $30\kms$, vs. total mass, $M_{\rm out}$,
    within the volume considered by \citet{Garrison-Kimmel2014b} in
    their study of the ``too-big-to-fail'' problem in the LG
    field. The volume consists of two spheres, each of radius $1.2$
    Mpc, centred on the two primaries but excluding the inner $300$
    kpc of each. Note that, for the same volume, the total number of
    massive objects decreases by a factor of $\sim 1.5$ from \apdmo
    (dark matter only; open circles) to \aphydro (filled circles), as
    noted by \citet{Sawala2013,Sawala2015}. The grey band spans the
    number of massive haloes in the ELVIS DMO simulations of
    \citet{Garrison-Kimmel2014}.}
\label{FigTBTF}
\end{figure}

\subsection{Implications for the too-big-to-fail problem}
\label{sec:tbtf}

The results from the previous section have implications for the
``too-big-to-fail (TBTF) problem in the field'' raised by
\citet[][hereafter GK14]{Garrison-Kimmel2014b}. This problem concerns
the number of haloes in DMO simulations that are too massive to be
consistent with any observed galaxy with robust kinematic and
photometric measurements. \citetalias{Garrison-Kimmel2014b} considered
systems with $V_{\rm peak} > 30 \kms$, where $V_{\rm peak}$ is the
maximum value of $V_{\rm max}$ attained by a system throughout its
history.

The counts of such objects in a volume defined by the combination of
two spheres of radius $1.2$ Mpc centred on each primary, but excluding
their inner $300$ kpc (which are populated by satellites), is reported
to be in the range $\sim22$-$48$ in the
\citetalias{Garrison-Kimmel2014b} ``ELVIS'' simulations\footnote{These
  values are estimated from figure 6 of GK14 and can change by
  1-2.}. These were contrasted with the $\sim 10$ known galaxies with
kinematic measurements consistent with $V_{\rm peak}>30 \kms$ haloes.
The difference between the two is the basis for the TBTF problem in the field.

We revisit this issue in \fig{FigTBTF}, where the open circles
indicate the number of such haloes in the $12$ AP-DMO simulations (L2),
plotted as a function of the total mass, $M_{\rm out}$, in the same
``hollowed-out spheres'' used by
\citetalias{Garrison-Kimmel2014b}. As expected given our discussion in
the previous subsection, the total number of haloes correlates
strongly with $M_{\rm out}$, which, in turn, correlates strongly with
$M_{\rm 3Mpc}$. As in \fig{FigDove2}, the numbers are systematically
reduced in the \aphydro runs, due to the effect on $V_{\rm max}$
caused by the loss of baryons.

As the figure shows, half of all \aphydro volumes have fewer than $22$
systems with $V_{\rm peak}>30 \kms$ and two of them have 10 or fewer,
reducing substantially the number of haloes without observed
counterparts (``massive failures'', in the parlance of
\citetalias{Garrison-Kimmel2014b}). The TBTF problem would only be
manifest in volumes with total masses well in excess of $M_{\rm
  3Mpc}\sim 10^{13}\Msun$, which, as we shall discuss below, are
disfavoured on other grounds.

\red{This constraint on $\Mout$ implies that the total virial mass of
  primaries should be less than $M_1+M_2\sim 3\times10^{12} \Msun$
  (\fig{FigDove1}). This is consistent with recent estimates of
  $M_1+M_2$ based on the LG kinematics \citep{Fattahi2016}, and of the
  MW virial mass, $M_{200}\sim10^{12}\Msun$ \citep[see; e.g.,][ and
  references therein]{Callingham2019}, if M31's virial mass does not
  exceed the Milky Way's by more than a factor of $\sim 2$, as seems likely \citep{Penarrubia2016}.}

\begin{figure*}
  \hspace{-0.2cm}
  \resizebox{17cm}{!}{\includegraphics{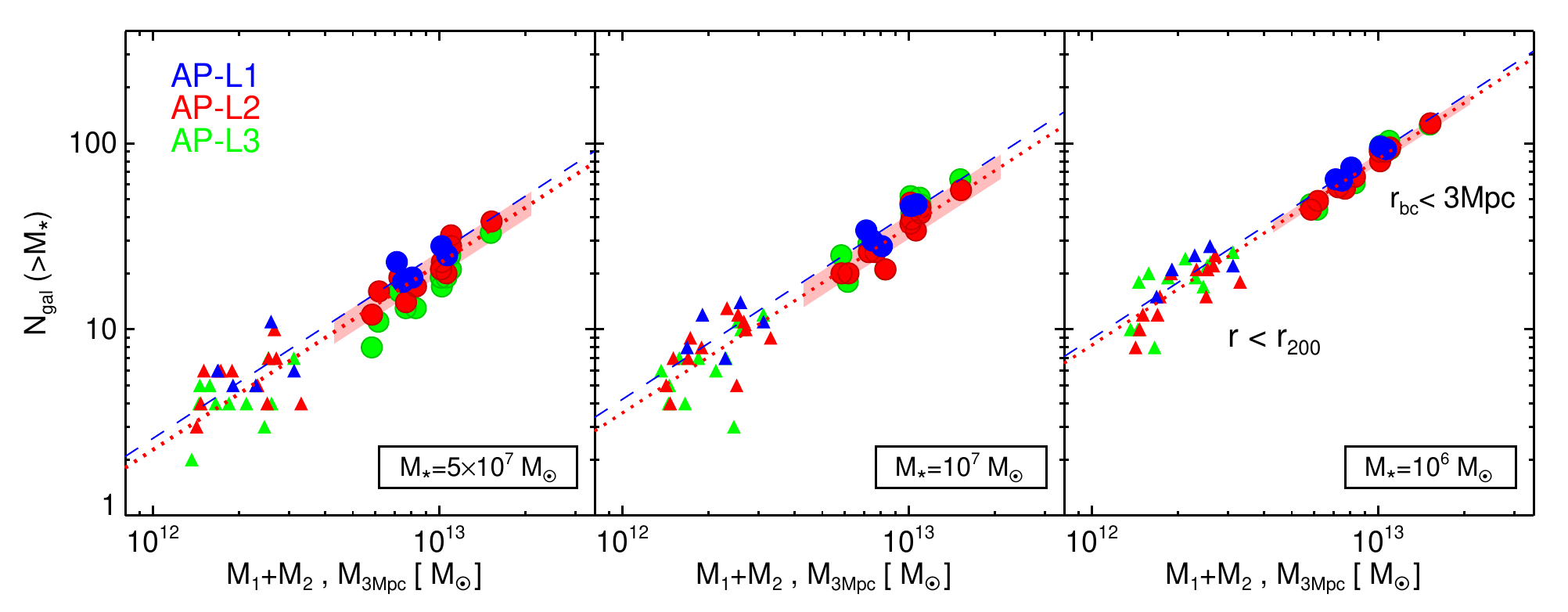}}\\%
  \caption{Number of galaxies above a given stellar mass, as specified in the
    legends, versus either $M_{\rm 3Mpc}$ (circles) or the combined virial mass
    of the primaries, $M_1+M_2$ (triangles) for \aphydro simulations
    of varying resolution. Dotted and dashed
    lines are power-law fits with unity slope  ($N\propto M$) to the
    AP-L1 and AP-L2 filled circles. The light red shaded
    regions represent the rms scatter around the AP-L2 fit. Note that
    the same fit equally well both circles and triangles, indicating
    that the dwarf galaxy formation efficiency in and around the
    primaries is very similar.}
\label{FigNgalMass}
\end{figure*}

\begin{figure}
  \hspace{-0.2cm}
  \resizebox{8.cm}{!}{\includegraphics{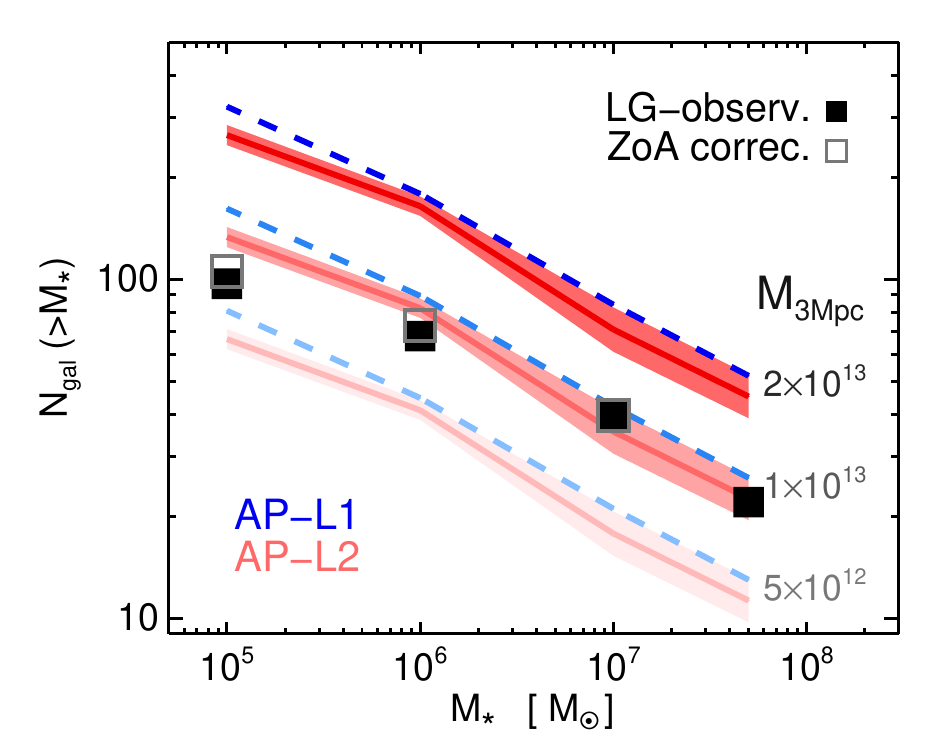}}\\%
  \caption{Predicted number of galaxies brighter than a given stellar
    mass from the fits to \aphydro in Fig.~\ref{FigNgalMass}, for
    different values of $M_{\rm 3Mpc}$. Blue (dashed) and red (solid)
    lines correspond to results for AP-L1 and AP-L2 fits,
    respectively. The observed number of galaxies in the Local Group
    are shown by the solid squares. Open squares are the same,
    corrected for galaxies possibly missing in the zone of avoidance
    (see text for details).}
\label{FigLF1}
\end{figure}

\subsection{Galaxy formation efficiency}
\label{sec:galeff}

A similar trend with total LG mass to that described for haloes in
Sec.~\ref{sec:subhaloes} is seen for the number of simulated galaxies
above a certain value of $M_*$. This is shown in \fig{FigNgalMass},
where we plot results for simulated galaxies with stellar mass
exceeding $10^6$, $10^7$ and $5\times 10^7 \Msun$. The linear trend is
remarkably tight in all cases, and extends all the way to the
satellites of individual primaries. This suggests that the tidal
stripping effects that reduce the numbers of subhaloes above a fixed
value of $V_{\rm max}$ (i.e., triangles in Fig.~\ref{FigDove2}) are
less important when considering stellar mass. This is not unexpected,
as stars are confined to the bottom of the subhalo potential well,
where they are harder to strip than the surrounding dark
matter. Subhaloes can thus lose substantial amounts of dark matter
before their stellar content is significantly altered.

The main lesson from \fig{FigNgalMass} is that, at least in this mass
range, the ``efficiency'' of dwarf galaxy formation in or around the
primaries of our LG simulations is remarkably similar. \red{In other
  words, the number of dwarfs per unit mass is independent of whether
  the count is carried out over the virialized region of the primaries
  or over the whole LG volume.} This implies that the fraction of all
LG galaxies above a certain $M_*$ that are satellites of either the MW
or M31 should be approximately the same as the fraction of the total
mass within $3$ Mpc that is contained within the virial boundaries of
the two primaries.

\red{Note that this conclusion is insensitive to our adoption of the
 primary's virial radius to identify ``satellites''. Indeed, changing the
  definition from $r_{200}$ to a fixed radius of $300$ kpc, for example, would redefine
  primary masses and satellite numbers in similar proportion, simply
  shifting systems along the $1$:$1$ line in
  \fig{FigNgalMass}. Although we do not show it here, we have
  explicitly checked that this is the case.}

For example, there are $24$ galaxies with $M_*>5\times 10^7 \Msun$ in
the LG volume, 8 of which are satellites. The satellite fraction is
thus $\sim 30$ per cent, consistent with the $\sim 1$:$4$ mass ratios
inferred from \fig{FigDove1}. For $M_*>10^7 \Msun$ the satellite
fraction is also similar, with $12$ satellites out of a total of $42$
galaxies. These numbers seem to validate our simulation result that
$M_{\rm 3Mpc}$ should be roughly $3$-$4$ times the combined virial
mass of the primaries. Note that this conclusion is independent of the
actual virial mass of the primaries, which are not accurately known
\citep[see,][and references therein]{Callingham2019}.

\subsection{The total mass of the Local Group}
\label{sec:LGmass}

The results of the previous subsection imply that, properly
calibrated, the total number of dwarfs may be used as a proxy for the
total mass, subject to an appropriate normalisation. For example, the
raw numbers of observed galaxies with $M_*>5\times 10^7$ or
$>10^7\Msun$ may be used to infer the total mass of the Local Group,
using the \apostle galaxy mass-halo mass relation, which is
responsible for the vertical normalisation of the lines in
\fig{FigNgalMass}.

Coloured lines in \fig{FigLF1} show the cumulative number of \apostle
galaxies as a function of stellar mass for different values of
$M_{\rm 3Mpc}$: $5\times 10^{12}$, $10^{13}$, and
$2\times 10^{13}\Msun$, respectively. The line colours indicate AP-L2
(red) and AP-L1 (blue), and the shaded area is the 1-$\sigma$ scatter
expected from the normalisation uncertainty for AP-L2 runs. The
observed numbers of galaxies (solid squares) with $M_*>10^7\Msun$
clearly favour $M_{\rm 3Mpc}\sim 10^{13}\Msun$, with a statistical
uncertainty much smaller than a factor of $2$. In particular, fitting
AP-L2 results to galaxies with $M_*>5\times 10^7$ or $10^7\Msun$ yield
$M_{\rm 3Mpc}=(1.0 \pm 0.15 )\times 10^{13}$ and
$(1.18\pm 0.17)\times 10^{13}\Msun$ for each case, respectively. We
emphasise that these estimates are sensitive to the \apostle
normalisation of the galaxy formation efficiency, which may vary for
other models of star formation, feedback, and reionisation.

We note that the estimated mass is sensitive to the minimum dwarf
galaxy stellar mass
chosen to match the curves in \fig{FigLF1}.  Indeed, taken at face
value, the total number of galaxies above $M_*=10^5\Msun$ would
suggest almost a factor of two lower $M_{\rm 3Mpc}$. This implies that
either the galaxy formation efficiency varies substantially with
stellar mass in and around the LG primaries, or that our current LG
inventory is missing about $\sim 50$ dwarfs at least as massive as
$M_*=10^5\Msun$.  We favour the latter explanation, not only because
the former is at odds with the simulation results, but also because,
as we discuss below, the evidence for incompleteness in the LG
inventory below $10^7\, M_\odot$ is quite compelling.

\begin{figure}
  \hspace{-0.2cm}
  \resizebox{8.cm}{!}{\includegraphics{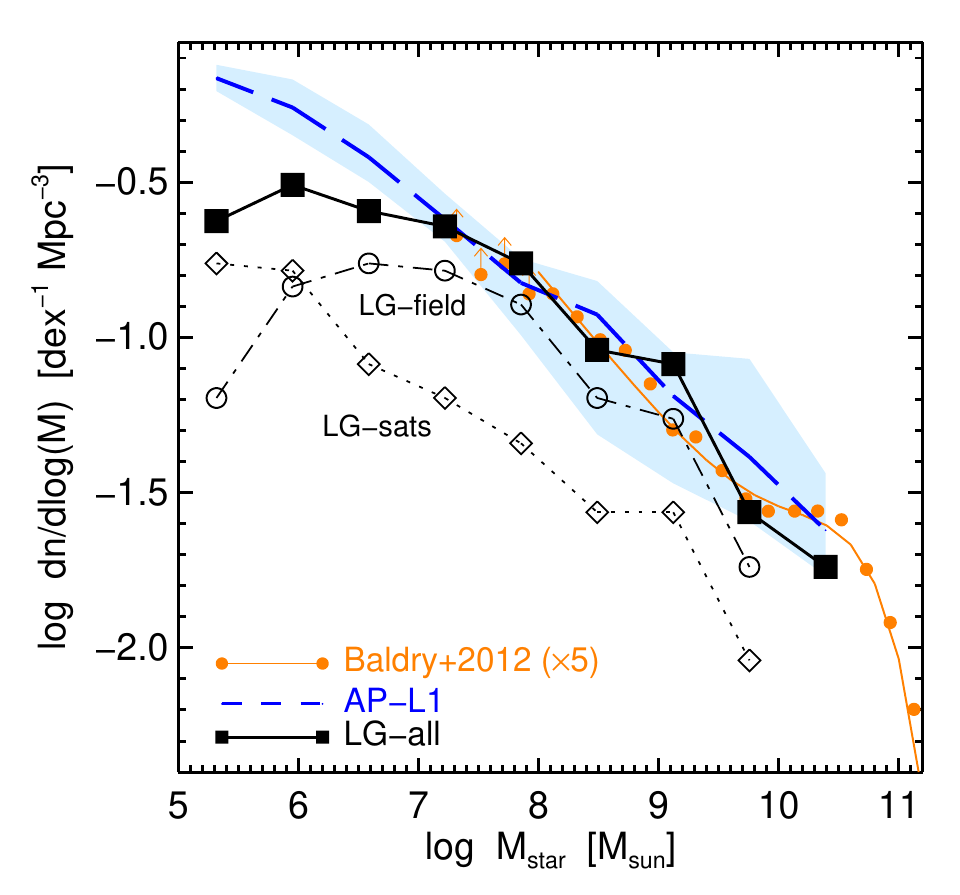}}\\%
  \caption{The observed stellar mass function of LG galaxies within
    $3$ Mpc from the LG barycentre. Solid squares correspond to all
    known galaxies in that volume, while small circles and diamonds
    illustrate the contribution from field galaxies and from MW/M31
    satellites. The average stellar mass function of $5$ AP-L1 runs,
    each normalised to a common $M_{\rm 3Mpc}=10^{13} \Msun$ is shown
    by the solid blue line. The shaded blue region spans the full
    range of variation of individual runs. The field galaxy stellar
    mass function from \citet{Baldry2012} ($\times 5$), together with
    their double Schechter fit, are shown in orange.}
  \label{FigLF}
\end{figure}

\subsection{Incompleteness in the LG inventory}
\label{SecIncompleteness}

We begin by noting that the incompleteness suggested by our discussion
above is much larger than what may be expected purely from the highly
extincted ``zone of avoidance'' delineated by the Galactic disk. This
area obscures a region of roughly $\pm 15$ degrees around the Galactic
plane, which translates into $\sim 30$ per cent of the available
sky. Correcting for this effect\footnote{\red{The correction is
    estimated by assuming the number density of galaxies inside the
    zone of avoidance is similar to the number density outside it.}}
would only lift the number of $M_*>10^5\Msun$ field dwarfs by $\sim
8$, as shown by the open symbols in \fig{FigLF1}. This is much smaller
than is required to bring the number of $M_*>10^5\Msun$ dwarfs into
agreement with what is expected for $M_{\rm 3Mpc}=10^{13}\Msun$.

Compelling evidence for incompleteness comes from the observed LG
galaxy stellar mass function, shown by the solid squares in
\fig{FigLF}. For $M_*>10^7\Msun$, the shape of this function compares
well with that of \citet[][suitably scaled\footnote{The scaling shown
    in Fig.~\ref{FigLF} is different from the one expected from the
    $M_{\rm 3Mpc}$ estimates of Sec.~\ref{sec:LGmass}. This is because
    the \apostle galaxy stellar mass function, like that of the EAGLE
    simulation, does not match perfectly the \citet{Baldry2012}
    results. We refer the reader to \citet{Schaye2015} for further
    details on this issue.}]{Baldry2012}, as well as with the AP-L1
galaxy mass function (blue curve and shaded region), normalised to
$M_{\rm 3Mpc}=10^{13}\Msun$. Below $10^7\Msun$, however, the observed
LG mass function flattens and becomes much shallower than for
\apostle. This flattening of the mass function below $10^7\Msun$ is a
sign of incompleteness and is the reason behind the deficit of dwarf
galaxies at $10^5\Msun$ highlighted when discussing \fig{FigLF1}.

The open circles (corresponding to observed LG field galaxies) and
diamonds (satellites) show that the flattening is entirely driven by a
pronounced lack of low-mass galaxies in the field: the MW and M31
satellites have a steeper faint end slope, actually consistent in
shape with that of the \apostle mass function. \red{This is in
  agreement with \citet[e.g.][]{Newton2018}, who have argued that the
  luminosity function of MW satellites is almost complete above
  $M_* \gtrsim 10^5\Msun$ (i.e. $M_{\rm V} \lesssim -8)$.}

The flattening in the LG stellar mass function below $10^7\Msun$ is
thus most likely the result of incompleteness in existing catalogues,
as satellites are easier to find than LG field dwarfs. Indeed, they
are closer to the Sun in the case of the Milky Way, and they are
concentrated in a smaller region of the sky in the case of M31, which
makes them easier to survey to faint levels \citep[e.g., Pan-Andromeda
Archaeological Survey, PAndAs,][]{Martin2006,McConnachie2009}.

\subsection{Missing dwarfs in the Local Group}
\label{sec:missingDwarfs}

The discussion above implies that $\sim 50$ dwarf galaxies with
stellar mass greater than $10^5 \Msun$ are missing from the current LG
inventory. This prediction depends solely on assuming that the
efficiency of dwarf galaxy formation is uniform throughout the volume
(as found in the simulations) and that the inventory of the most
luminous dwarfs is complete. Or, in other words, these are the number
of missing dwarfs in the field required to make the satellite fraction
of $M_*>10^5\, M_\odot$ dwarfs the same as that of galaxies with
$M_*>10^7\, M_\odot$.

\begin{figure}
  \hspace{-0.2cm}
  \resizebox{8.cm}{!}{\includegraphics{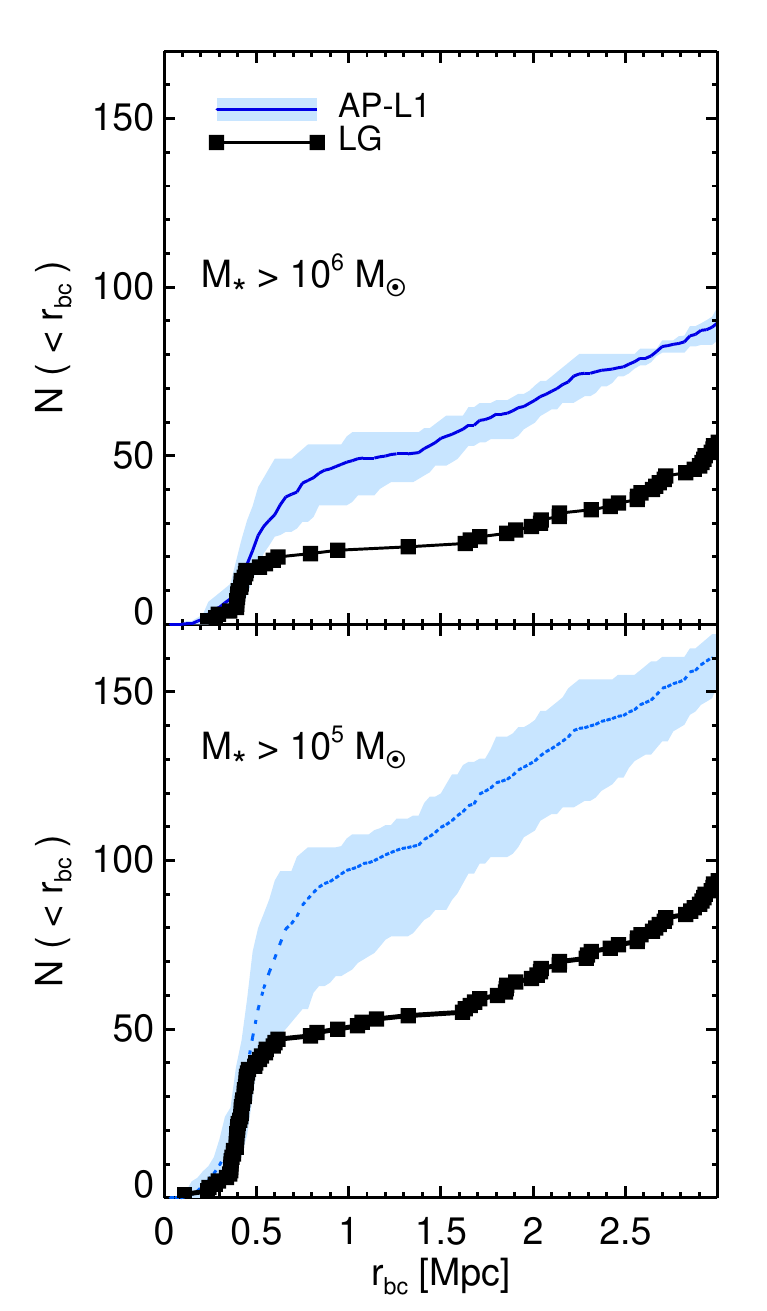}}\\%
  \caption{{\it Top}: The cumulative radial distribution of observed LG
    galaxies with stellar mass above $10^5 \Msun$, shown as connected
    filled squares. Radii are measured from the LG barycentre. The
    solid blue line indicates the results expected from AP-L1 runs,
    normalised to a common mass of $M_{\rm 3Mpc}=10^{13}\Msun$. The
    shaded region illustrate the full range of variations for
    individual AP-L1 runs. {\it Bottom}: Same as top panel but for the
    stellar mass cut of $\Mstr > 10^6 \Msun$.}
  \label{FigRdist}
\end{figure}

Where are these missing galaxies located? Naively, one might
anticipate that they are predominantly near the outer edge of the
$3$~Mpc sphere, where much of the volume resides. The simulations, however,
suggest otherwise. We show this in \fig{FigRdist}, where the blue
curve in the top panel shows the average cumulative
radial distribution of galaxies more massive than $10^6 \Msun$  (plus scatter) in
AP-L1 volumes, measured from their barycentre, after normalising them
to a common mass of $M_{\rm 3Mpc}=1 \times 10^{13}\Msun$; the bottom
panel shows the results for simulated galaxies more massive than $10^5
\Msun$.

\begin{figure}
  \hspace{-0.2cm}
  \resizebox{8.cm}{!}{\includegraphics{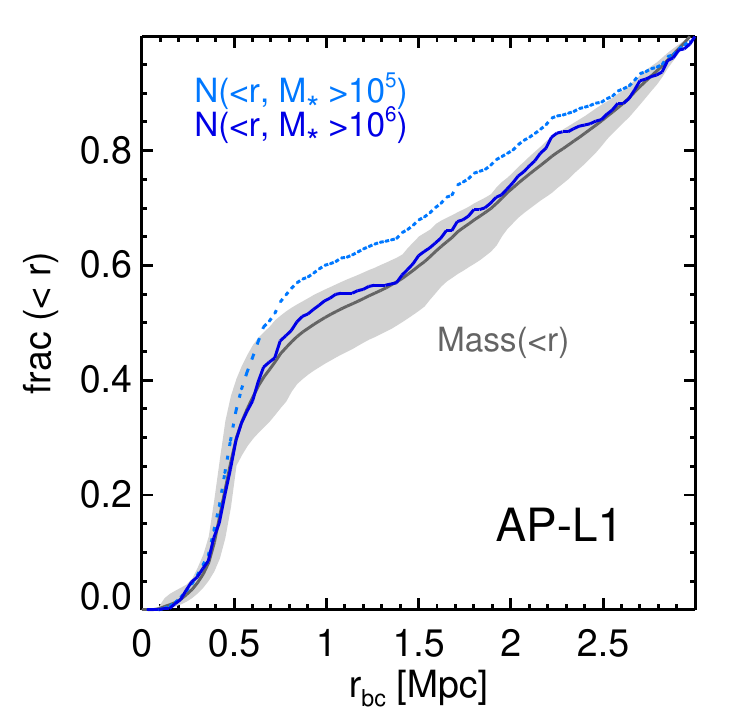}}\\%
  \caption{Average cumulative mass profile of AP-L1 runs, normalised
    to their $M_{\rm 3Mpc}$ (in grey). Curves indicate cumulative
    numbers of galaxies with $M_*>10^5\Msun$ (cyan) or
    $M_*>10^6\Msun$ (blue), respectively. Note that the number of
    galaxies traces closely the mass distribution in the LG
    simulations.}
  \label{FigRdistMass}
\end{figure}

According to the \apostle results, nearly half of all galaxies are
expected to be within the inner $1$ Mpc (i.e., in the inner $\sim 4$
per cent of the full volume), and $75$ per cent are expected to be
within $2$ Mpc, occupying only the inner $\sim 30$ per cent of the
volume. As anticipated by our earlier discussion on galaxy formation
efficiency, \fig{FigRdistMass} shows that the cumulative radial
distribution of galaxies neatly tracks the distribution of mass within
the \apostle volumes.

\begin{figure*}
  \hspace{-0.2cm}
  \resizebox{17.5cm}{!}{\includegraphics{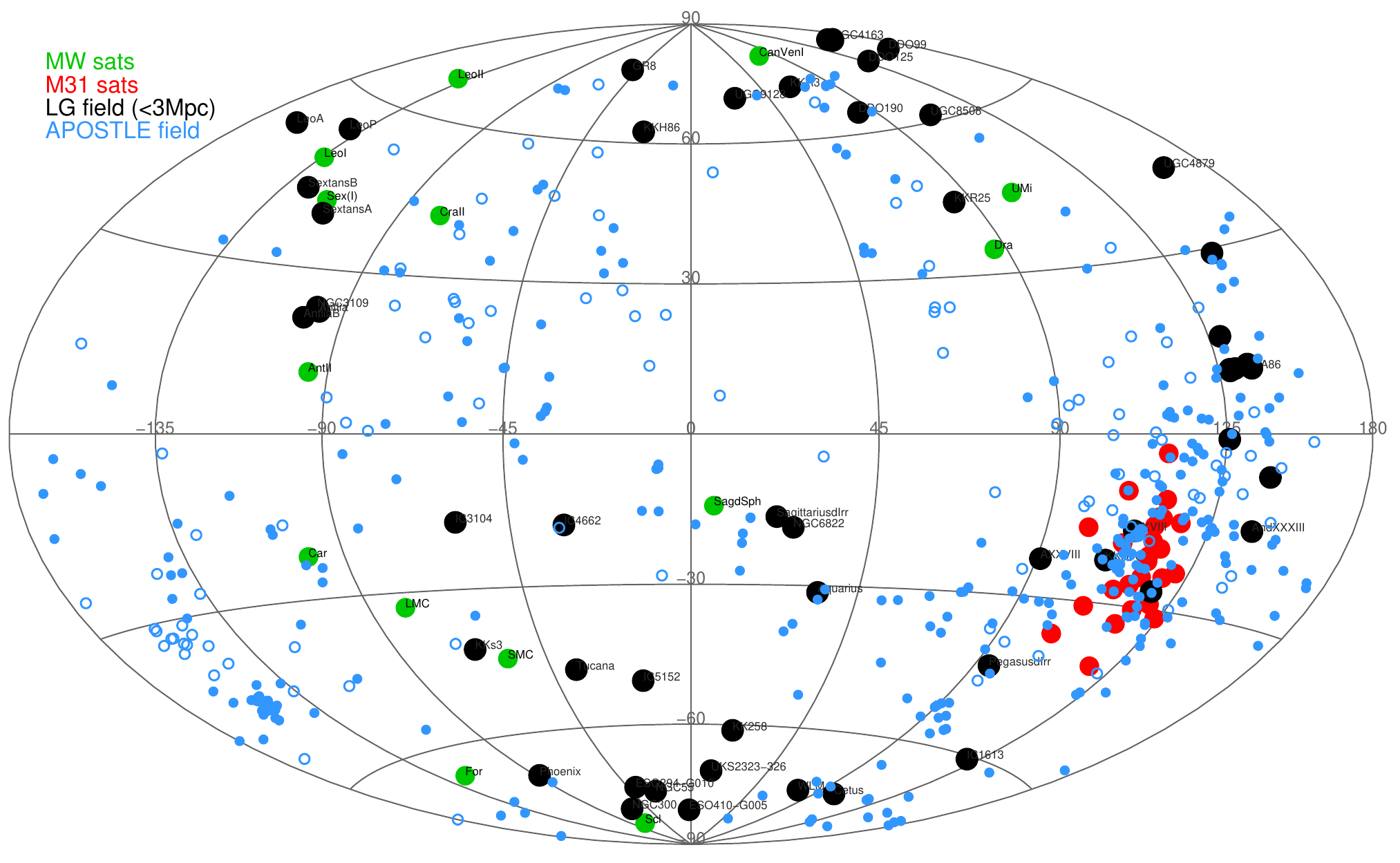}}\\%
  \caption{Aitoff projection of observed LG dwarf galaxies with
    stellar mass above $10^5\Msun$, in Galactic coordinates. MW and
    M31 satellites are shown with green and red symbols,
    respectively. LG field dwarfs (i.e., within 3Mpc from the LG
    barycentre) are shown with large black circles. Simulated {\it
      field} dwarf galaxies with $M_*>10^5 \Msun$ in $5$ AP-L1
    simulations are shown with small blue circles. Open and field
    symbols indicate whether the simulated dwarf is closer or farther
    than $1$ Mpc from the MW analogue, respectively. The simulation
    coordinates are centred on the MW analogues and rotated so that
    the M31 analogues are in the same sky location as M31; i.e.,
    $(l,b)=(121.2^{\circ},-21.6^{\circ})$. Note that many missing
    dwarfs are expected to be in the region that immediately surrounds
    M31, but most are expected to be more than $1$ Mpc away, and thus
    may have been missed by existing surveys.}
  \label{FigSky}
\end{figure*}

\begin{figure}
  \hspace{-0.2cm}
  \resizebox{8.7cm}{!}{\includegraphics{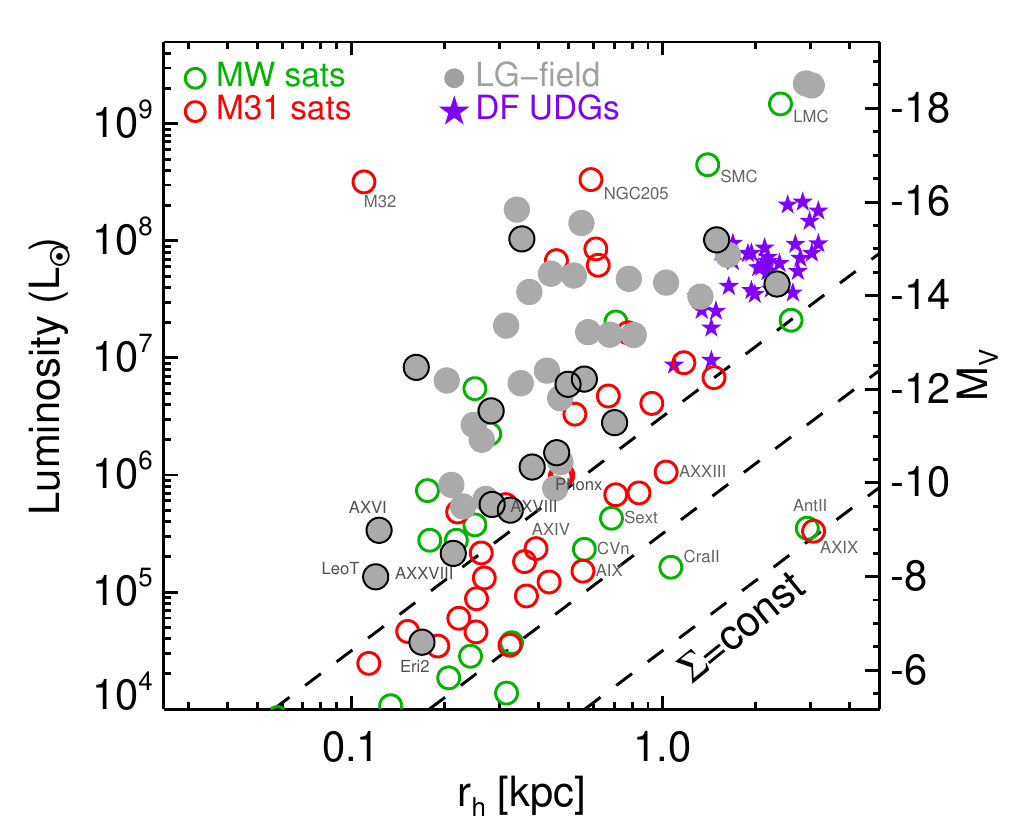}}\\%
  \caption{Luminosity vs. half-light radii of observed LG field and satellite
    dwarf galaxies, shown as solid squares and open circles,
    respectively. Field galaxies with $\sigmalos$ measurements are
    highlighted with black squares. For comparison, Dragonfly UDGs of
    the Coma cluster \citep{vanDokkum2015} are shown as green starred
    symbols. We use r-band luminosities for these galaxies. The
    diagonal lines illustrate lines of constant effective surface
    brightness, each separated by 1-dex and corresponding to
    $\Sigma_{\rm eff}=10^6$, $10^5$, and $10^4 \, {\rm
      L_\odot}$/kpc$^2$, equivalent to roughly $26.5$, $29$, and
    $31.5$ m$_V$/arcsec$^2$. }
  \label{FigMassR}
\end{figure}

Indeed, most of the ``missing dwarfs'' are expected to be within $1.5$
Mpc of the LG barycentre, mainly clustered in the outskirts of the
haloes of the two primaries. This is illustrated in \fig{FigSky},
which shows, in an Aitoff all-sky projection and viewed from the MW
perspective, the location of all AP-L1 field galaxies with $M_*>10^5
\Msun$ (all $5$ AP-L1 volumes combined; see small blue circles). The
coordinate system is defined so that the M31-analogues are located in
the same sky position as in observations. Observed field dwarfs are
presented as black circles, and MW and M31 satellites as green and red
ones, respectively. Interestingly, \apostle predicts that (missing)
field dwarf galaxies are {\it not} randomly located on the sky, but
more likely towards M31's direction\footnote{\red{This is consistent
    with observations; \citet{McConnachie2012} points out that known
    field dwarf galaxies in the LG are located preferentially towards
    M31 in the sky.}}, with some of them being closer than 1~Mpc to
the MW (open circles).

Why have these dwarfs been missed? The most likely explanation is that
they are extended, low surface brightness systems that do not stand
out in panoramic surveys. This is easily appreciated in
\fig{FigMassR}, where we show the $V$-band luminosity and stellar
half-mass radius of all LG dwarfs, if available. MW and M31 satellites
are shown by open circles, whereas LG field dwarfs are indicated by
filled squares. Dashed lines indicate constant effective surface
brightness, each separated by one dex, and starting, at the top, at
$10^6 \,{\rm L_{\odot}}/$kpc$^2$ (i.e., $\sim 26.5$
m$_V$/arcsec$^2$).

Note that this is already below the effective surface brightness of
``ultra-diffuse'' galaxies (UDGs) such as the Coma cluster Dragonfly
galaxies \citep[DF, shown for reference with starred green
symbols,][]{vanDokkum2015}. Clearly, many observed LG galaxies are
extremely low surface brightness systems far fainter than typical
UDGs. Such LG galaxies are typically resolved into individual stars
and their discovery relies on special methods based on searching for
overdensities after filtering stars with isochrone masks \citep[see,
e.g.,][]{Koposov2008}.

The extreme LSB regime probed by LG dwarfs reaches, in the case of
And~XIX or Crater~II, approximately $\Sigma_{\rm eff} \sim 30$
m$_V$/arcsec$^2$. Indeed, $\sim 40$ per cent of all known LG
satellites with $M_*>10^5 \Msun$ have effective surface brightness
below the Coma UDGs. By contrast, only {\it one} LG field galaxy,
Eri~II, has an effective surface brightness below
$10^6 \,{\rm L_\odot}/$kpc$^2$. Eri~II was discovered relatively
recently and is located just outside 300 kpc from the MW
\citep{Li2017}.

\red{\fig{FigMassR} also suggests that, in the range
$10^5<M_*/M_\odot<10^6$, the properties of the missing galaxies should
be similar to those of M31 satellites such as And XIV, And XIX, or And
XXIII, which were identified in the PAndAS survey at a distance of
$\sim 700$ kpc (We emphasize that these are {\it not} ``ultra-faint''
dwarfs, but, rather, systems with luminosities comparable to that of
``classical'' dSphs). Putting all these results together, we conclude
that a PAndAS-like survey of the outskirts of the M31-M33 system, if
deep enough to detect systems like the aforementioned M31 satellites
out to $1.5$-$2$ Mpc, should be able to net the majority of the
isolated dwarfs missing from our current inventory of the Local Group.
We note that at a distance of $1.5$ Mpc, even stars as
bright as blue horizontal branch stars would have magnitudes
$m_V\sim 26$.}

\section{Summary}
\label{sec:summary}

We have analysed the environment of galaxy pairs with mass and
kinematics resembling the Milky Way and Andromeda galaxies in the
\dove N-body simulation of a large cosmological volume. We find that
the total mass within $3$ Mpc from the pair midpoint (which we define
as the Local Group boundary) is typically $\sim 4$ times the combined
virial mass of the pair. In general, within that volume there is
always at least as much mass within the virial boundaries of the
primaries as in their surroundings.

Cosmological hydrodynamical re-simulations of many of those pairs
(from the \apostle Project) show that the dwarf galaxy formation
efficiency, defined as the total number of dwarfs above a certain
stellar mass per unit mass, is uniform throughout the volume and thus
similar in and around the primaries. This implies that the total
number of dwarfs within $3$ Mpc, or within the virial volume of each
primary, depends solely on the total mass of that volume.

These two results indicate that satellites should make up about $1/4$
of all dwarf galaxies in the Local Group. Although this prediction is
approximately correct for observed dwarfs more massive than $M_*=10^7\Msun$,
the agreement becomes gradually worse for less massive galaxies:
satellites make up more than $40$ per cent of those with $M_*>10^5\Msun$.
We interpret this as a result of incompleteness in existing LG dwarf
galaxy catalogues, and predict that there are $\sim 50$ dwarfs at
least as massive as the Draco dSph (i.e., $M_*\sim 10^5\Msun$)
missing from our current LG inventory.

Our interpretation is supported by (i) the faint-end shape of the LG
field galaxy luminosity function, which, unlike that of satellites,
becomes abruptly shallow below $M_*\sim 10^7\Msun$, and by (ii)
the lack of LG field galaxies with effective surface brightness below
$\sim 10^6 \,{\rm L_\odot}$/kpc$^2$.

The simulations indicate that the missing dwarfs should cluster
tightly around the two primaries. Indeed, most of them should be
within $1.5$ Mpc from the LG barycentre. Additionally, a notable
fraction is expected to lie around M31 in the sky. These are strong
predictions that might be possible to verify with upcoming wide-field
deep imaging surveys, such as that planned by the Large Synoptic
Survey Telescope or \red{the Canada-France Imaging Survey}, if
existing methods for detecting dwarfs using ground-based imaging can
be improved enough to target such objects, or if planned wide-field
imaging surveys from space \citep[such as that envisioned by the
  MESSIER or CASTOR
  missions\footnote{\url{http://orca.phys.uvic.ca/~pcote/castor/}},][]{Messier2017,Castor2012}
come to fruition.

If proven correct, our results have strong implications for a number
of current discussions regarding perceived failures of the
$\Lambda$CDM paradigm in the Local Group. Issues such as the
``too-big-to-fail'' problem in the field \citep{Garrison-Kimmel2014b},
where the kinematics of observed dwarfs are used to infer their halo
masses, whose abundance is, in turn, compared with simulations, can be
singularly affected. In particular, if the total mass of the Local
Group is about $M_{\rm 3Mpc}\sim 10^{13}\, M_\odot$, the total number
of massive haloes without observed counterparts is not significant,
especially when considering the reduction in mass that low-mass haloes
experience because of the loss of baryons as a result of reionisation
and feedback \citep{Sawala2013,Sawala2015}.

Finally, the fraction of known dwarfs in the Local Group volume
with kinematic measurements is relatively small (only $\sim 30$ per cent of
LG field dwarfs have published velocity dispersion), and our results
raise the possibility that many existing dwarfs may actually be
missing from current catalogues. As the Local Group inventory of dwarf
galaxies becomes increasingly complete, it is likely that our
understanding of dwarf galaxy formation in our immediate cosmic
neighbourhood, and its cosmological significance, will become clearer.

\section{Acknowledgements}

We are thankful to Matthieu Schaller, Kyle Oman and Till Sawala for
reviewing the manuscript, as well as Nicolas Martin for fruitful
discussions. We thank the anonymous referee for their comments. AF is
supported by a EU COFUND/Durham Junior Research Fellowship (under EU
grant agreement no. 609412) and the STFC grant ST/P000541/1. CSF
acknowledges support by the European Research Council (ERC) through
Advanced Investigator grant DMIDAS (GA 786910).

This work greatly benefited from the DiRAC Data Centric system at
Durham University, operated by the ICC on behalf of the STFC DiRAC HPC
Facility (\href{https://dirac.ac.uk/}{www.dirac.ac.uk}). This
equipment was funded by BIS National E-infrastructure capital grant
ST/K00042X/1, STFC capital grant ST/H008519/1, and STFC DiRAC
Operations grant ST/K003267/1 and Durham University. DiRAC is part of
the National E-Infrastructure. This work party used the computing and
storage hardware provided by WestGrid (www.westgrid.ca) and Compute
Canada Calcul Canada (www.computecanada.ca), as well as UK Research
Data Facility (http://www.archer.ac.uk/documentation/rdf-guide).


\bibliographystyle{mnras}
\bibliography{master}
\bsp

\label{lastpage}
\end{document}